\documentclass[%
pdflatex,
reprint,
superscriptaddress,
frontmatterverbose,
preprintnumbers,
nobibnotes,
nofootinbib,
amsmath,amssymb,
aps,
prl
]{revtex4-2}
\usepackage[pdftex]{graphicx}
\usepackage{dcolumn}
\usepackage{bm}
\usepackage{CJK}
\usepackage[T1]{fontenc}
\usepackage{multirow}

\def\ve#1{{\bm{#1}}}
\def\nuc#1#2#3{{}^{#2}_{#3}\mathrm{#1}}
\def\urm#1{\scriptstyle{\text{\textrm{\textmd{\textup{#1}}}}}}

\def\avr#1{\left\langle{#1}\right\rangle}
\begin{document}
% 
%%%%%%%%%%%%%%%%%%%%%%%%%%%%%%%%%%%%%%%%%%%%%%%%%% 
\begin{CJK*}{UTF8}{}
  \preprint{RIKEN-iTHEMS-Report-23}
  \title{QCD-based charge symmetry breaking interaction \\
    and the Okamoto-Nolen-Schiffer anomaly}
  \author{Hiroyuki Sagawa (\CJKfamily{min}{佐川弘幸})}
  \affiliation{
    RIKEN Nishina Center for Accelerator-based Science, Wako 351-0198, Japan}
  \affiliation{
    Center for Mathematics and Physics, University of Aizu, 
    Aizu-Wakamatsu, Fukushima 965-8560, Japan}
  \author{Tomoya Naito (\CJKfamily{min}{内藤智也})}
  \affiliation{
    RIKEN Interdisciplinary Theoretical and Mathematical Sciences Program (iTHEMS),
    Wako 351-0198, Japan}
  \affiliation{
    Department of Physics, Graduate School of Science, The University of Tokyo,
    Tokyo 113-0033, Japan}
  \author{Xavier Roca-Maza}
  \affiliation{
    Departament de F\'{\i}sica Qu\`{a}ntica i Astrof\'{\i}sica
    and
    Institut de Ci\`{e}ncies del Cosmos,
    Universitat de Barcelona, 
    Mart\'{\i} i Franqu\'{e}s 1,
    08028 Barcelona, Spain}
  \affiliation{
    Dipartimento di Fisica, Universit\`{a} degli Studi di Milano,
    Via Celoria 16, 20133 Milano, Italy}
  \affiliation{
    INFN, Sezione di Milano,
    Via Celoria 16, 20133 Milano, Italy}
  \author{Tetsuo Hatsuda (\CJKfamily{min}{初田哲男})}
  \affiliation{
    RIKEN Interdisciplinary Theoretical and Mathematical Sciences Program (iTHEMS),
    Wako 351-0198, Japan}
  \date{\today}
  %%%%%%%%%%%%%%%%%%%%%%%%%%%%%%%%%%%%%%%%%%%%%%%%%% 
  \begin{abstract}
    An approach is proposed to link the charge symmetry breaking (CSB) nuclear interaction and the low-energy constants in quantum chromodynamics (QCD) by matching the CSB effect in nuclear matter. 
    The resulting CSB interaction is applied 
    to study the Okamoto-Nolen-Schiffer
    anomaly, still lacking a satisfactory microscopic understanding, on the energy differences of mirror nuclei 
    by taking 
    $ \nuc{F}{17}{} $-$ \nuc{O}{17}{} $,
    $ \nuc{O}{15}{} $-$ \nuc{N}{15}{} $,
    $ \nuc{Sc}{41}{} $-$ \nuc{Ca}{41}{} $,
    and
    $ \nuc{Ca}{39}{} $-$ \nuc{K}{39}{} $ as typical examples.
    The magnitude and sign of the 
    QCD-based CSB interactions are found to 
    resolve the anomaly successfully within 
    theoretical uncertainties.
  \end{abstract}
  \maketitle
\end{CJK*}
%%%%%%%%%%%%%%%%%%%%%%%%%%%%%%%%%%%%%%%%%%%%%%%%%% 
% 
\par
An anomaly in the energy differences of mirror nuclei and isobaric analog states, not yet well understood from a microscopic point of view, was found more than 50 years ago and is called
the Okamoto-Nolen-Schiffer (ONS) anomaly~\cite{
  Okamoto1964Phys.Lett.11_150,
  Nolen1969Annu.Rev.Nucl.Sci.19_471}.
It was first reported by Okamoto for
the $ \nuc{He}{3}{} $-$ \nuc{H}{3}{} $ system, and
Nolen and Schiffer made a systematic study from light to heavy nuclei within the framework of the independent-particle model to
find that the theoretical values of the energy difference underestimate the 
experimental values
by  $ 3 $--$ 9 \, \% $.
Extra corrections such as
the finite proton size, the center-of-mass effect, the Thomas-Ehrman effect, 
the isospin impurity, the electromagnetic spin-orbit interaction, the proton-neutron mass difference in the kinetic energy, the core polarization effect, and the
vacuum polarization, altogether explain only about $ 1 \, \% $ of the discrepancy~\cite{
  Shlomo1978Rep.Prog.Phys.41_957}. 
\par
A possible remaining source to fill the gap
is the charge symmetry breaking (CSB) nuclear interaction~\cite{
  Okamoto1964Phys.Lett.11_150,
  Negele:1971oux,
  Blunden1987Phys.Lett.B196_295,
  Suzuki1992Nucl.Phys.A536_141,
  Miller:1990iz}.
Recently, phenomenological CSB interactions (often taken to be a
Skyrme-type contact interaction) have been introduced
to systematically calculate the isospin symmetry breaking effect on top of the Coulomb interaction; 
they provide successful results for describing the isobaric analog states,
the mass differences of iso-doublet and iso-triplet nuclei,
and also the double-$ \beta $ decays~\cite{
  Roca-Maza2018Phys.Rev.Lett.120_202501,
  Baczyk2018Phys.Lett.B778_178,
  Sagawa2019Eur.Phys.J.A55_227,
  Baczyk2019J.Phys.G46_03LT01,
  Naito2022Phys.Rev.C105_L021304,
  Naito2022Phys.Rev.C106_L061306,
  Naito:2023fnm}. However, both the magnitude and the sign of the parameters in phenomenological CSB interactions have not been well determined.
Meanwhile, microscopic calculations of observables sensitive to isospin symmetry breaking terms in the nuclear Hamiltonian have also become available~\cite{
  Novario2023Phys.Rev.Lett.130_032501}
although CSB effects have not been isolated in detail.
\par
The aim of this Letter is to provide a quantum chromodynamics (QCD)-based understanding of CSB by making a quantitative link between the Skyrme-type CSB interactions~\cite{
  Sagawa2019Eur.Phys.J.A55_227,
  Wiringa} 
and
the CSB effect due to the $ u $-$ d $ quark mass difference in QCD~\cite{
  Henley1989Phys.Rev.Lett.62_2586,
  Hatsuda1991Phys.Rev.Lett.66_2851,
  Saito:1994tq}.
First, we perform a matching of the phenomenological and QCD-based calculations on the binding-energy difference between the  neutron and the proton in an infinite nuclear matter
to constrain the sign and magnitude of the phenomenological
CSB interactions.
Then, the results are utilized to study 
the mass difference of mirror nuclei $ \Delta E $ of $ \left( N \pm 1, Z \right) $ and $ \left( N, Z \pm 1 \right) $ with the closed-shell core
($ A = N +Z = 16 $ and $ 40 $) based on the Hartree-Fock (HF) wave functions, aiming to see whether the ONS anomaly can be resolved microscopically.
These examples are chosen to suitably isolate CSB effects with respect to those originating from charge independence breaking (CIB) and, thus, robustly test our approach.
\par
Let us start with  the binding-energy
difference between the neutron and the proton $ \Delta_{np} \left( \rho \right) $ in infinite nuclear matter
($ N = Z $) with the baryon density $ \rho $,
as defined by a difference of the momentum independent part of the Lorentz-scalar self-energies.
In the leading order of the $ u $-$ d $ quark mass difference and the quantum electrodynamics (QED) effect,
an approximate formula has been obtained from the QCD sum rules (QSR)~\cite{
  Hatsuda1991Phys.Rev.Lett.66_2851}:
\begin{subequations}
  \label{eq:mass-pn}
  \begin{align}
    \Delta_{np} \left( \rho\right)
    & \simeq 
      C_1 G \left( \rho \right)
      -
      C_2,
      \label{eq:mass-pn_a} \\
    G \left( \rho \right)
    & =
      \left(
      \frac{\avr{\bar{q} q}}{\avr{\bar{q} q}_0}
      \right)^{1/3} .
      \label{eq:mass-pn_b}
  \end{align}
\end{subequations}
Here, $ \avr{\bar{q} q} $ and $ \avr{\bar{q} q}_0 $ are, respectively,
the isospin averaged in-medium and in-vacuum chiral condensate.
The coefficient $ C_1 $ is proportional to  the $ u $-$ d $ quark mass difference 
$ \delta m $,~\footnote{The renormalization group invariant mass difference reads
  $ \delta m \equiv m_d - m_u \simeq 3.6 \, \mathrm{MeV} $~\cite{
    FlavourLatticeAveragingGroupFLAG:2021npn}.}
through the isospin-breaking constant 
$ \gamma \equiv {\avr{\bar{d} d}}_0/{\avr{\bar{u} u}}_0 - 1 $ 
as $ C_1= - a \gamma $ with a positive numerical constant $ a $ determined by
the Borel QSR method~\cite{
  Hatsuda1991Phys.Rev.Lett.66_2851}.
On the other hand,
$ C_2 $ is a constant originating both from $ \delta m $ and the QED effect,
and is written as  $ C_2 =  C_1 - \Delta_{np} \left( 0 \right) $,
where experimental neutron-proton mass difference in the vacuum is denoted by
$ \Delta_{np} \left( 0 \right) = m_n - m_p \simeq 1.29 \, \mathrm{MeV} $.
Equation \eqref{eq:mass-pn} is valid at low density
$ \rho < \rho_0 = 0.17 \, \mathrm{fm}^{-3} $ where the dimension-3 chiral condensate gives a dominant contribution in the operator product expansion in QSR.
In the following, we take
$ C_1 = 5.24^{+2.48}_{-1.21} \, \mathrm{MeV} $,
where the central value is obtained from $ \gamma = - 7.8 \times 10^{-3} $~\cite{
  Hatsuda1991Phys.Rev.Lett.66_2851}
and the uncertainty is estimated from 
$ \gamma  = - \left(\text{$ 6 $--$ 11.5 $} \right) \times 10^{-3}$~\cite{Narison}.
Since the $ C_2 $-term is density independent, it is canceled out in the  following analysis.
\par
Equation \eqref{eq:mass-pn} implies that $ \Delta_{np} \left( \rho \right) $
tends to decrease in the nuclear medium associated with the partial restoration of chiral symmetry
$ G \left( \rho \right) < 1 $.
The in-medium chiral condensate in the leading order with the Fermi-motion correction 
has a universal form~\cite{
  Hayano:2008vn,
  Gubler2019Prog.Part.Nucl.Phys.106_1}
\begin{subequations}
  \label{eq:QCD-Q}
  \begin{align}
    \frac{\avr{\bar{q} q}}{\avr{\bar{q} q}_0}
    & \simeq
      1
      +
      k_1  \frac{\rho}{\rho_0}
      +
      k_2  \left( \frac{\rho}{\rho_0} \right)^{5/3},
      \label{eq:QCD-Q_a} \\
    k_1
    & =
      -
      \frac{\sigma_{\pi N}  \rho_0 }{f_{\pi}^2 m_{\pi}^2} < 0,
      \qquad
      k_2
      =
      -
      k_1
      \frac{3k_{\urm{F0}}^2}{10 m_N^2}
      >
      0,
      \label{eq:QCD-Q_b}
  \end{align}
\end{subequations}
where $ \sigma_{\pi N} $ is the $ \pi $-$ N $ sigma term,
$ m_{\pi} $ ($ m_N $) is the pion (nucleon) mass, and $ f_{\pi}$ is the pion decay constant.
The Fermi-momentum of the symmetric nuclear matter
at saturation is denoted by
$ k_{\urm{F0}} = \left( 3 \pi^2 \rho_0 / 2 \right)^{1/3} = 268 \, \mathrm{MeV} $.
Systematic calculations using  the in-medium chiral perturbation theory
shows that the full chiral corrections up to next-to-next-to leading order 
over Eq.~\eqref{eq:QCD-Q} is numerically 
small for  $ \rho < \rho_0 $~\cite{Goda2013Phys.Rev.C88_065204}
(see Fig.~S.1 in Supplemental Material~\cite{SM}).  
Alternative evaluation of the higher-order chiral corrections
with the $ \Delta $-excitation \cite{Kaiser2008Phys.Rev.C77_025204}
does not change this conclusion (see Fig.~S.2 in Supplemental Material~\cite{SM}).
We note however that the values of $ \sigma_{\pi N} $ have large uncertainty:
On the basis of the present values
of $ \sigma_{\pi N} $ from the scattering data and the lattice QCD data
($ N_f = 2 $ and $ 2 + 1 $) summarized in Fig.~47 of the FLAG Review 2021~\cite{
  FlavourLatticeAveragingGroupFLAG:2021npn},
we employ a conservative estimation, $ \sigma_{\pi N} = 45 \pm 15 \, \mathrm{MeV} $.
This value and the error happen to be similar to the old estimation in Ref.~\cite{Gasser:1990ce}.
Corresponding values of $ k_{\urm{$ 1 $, $ 2 $}} $ are summarized in Table~\ref{tab:tab-k}.
We note that the recent data from the pionic atoms~\cite{
  Nishi2023Nat.Phys.19_788}
indicate that   
$ \avr{\bar{q}q} / \avr{\bar{q}q}_0 \left( \rho = 0.58 \rho_0 \right) = 0.77 \pm 0.02 $
which is consistent with the value obtained from Eqs.~\eqref{eq:QCD-Q_a} and \eqref{eq:QCD-Q_b}.
\begin{table}[tb]
  \centering
  \caption{
    The parameters $ k_1 $ and $ k_2 $ in Eq.~\eqref{eq:QCD-Q}
    corresponding to the adopted $ \sigma_{\pi N} $ value with
    $ m_{\pi} = 135 \, \mathrm{MeV} $, $ m_N = 938 \, \mathrm{MeV} $, 
    and $ f_{\pi} = 92.4 \, \mathrm{MeV} $.}
  \label{tab:tab-k}
  \begin{ruledtabular}
    \begin{tabular}{c|cc}
      $ \sigma_{\pi N}$ ($ \mathrm{MeV} $) & $ k_1 $ & $ k_2 $  \\
      \hline
      $ 45 \pm 15 $ & $ -0.38 \pm 0.13 $ & $ 0.0093 \pm 0.0031 $  
    \end{tabular}
  \end{ruledtabular}
\end{table}
\par
\begin{figure}[tb]
  \centering
  \includegraphics[width=1.0\linewidth]{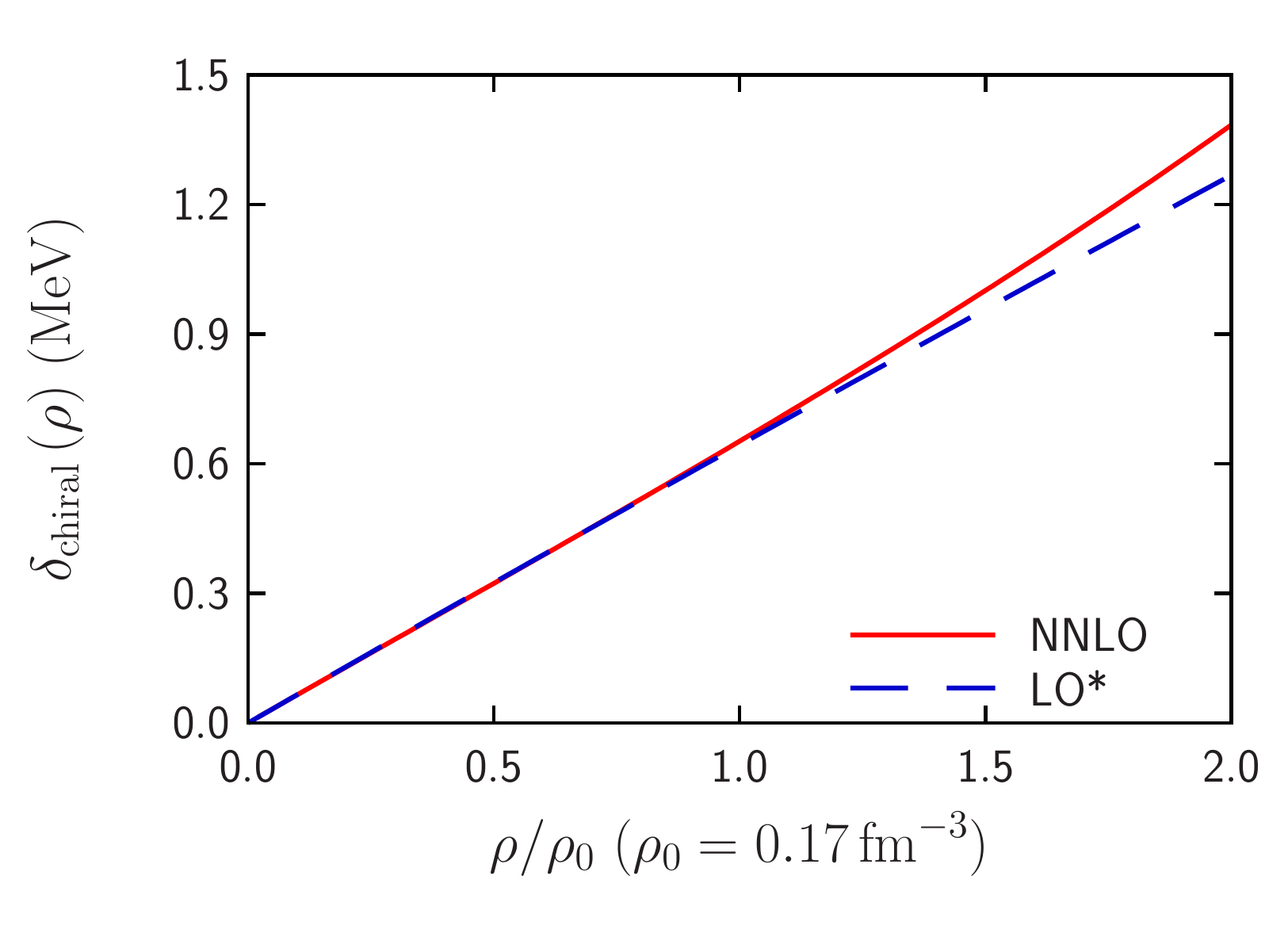}
  \caption{
    The CSB effect from the partial restoration of  chiral symmetry.  
    Blue dashed curve (LO*): the leading-order formula with Fermi-motion correction 
    [Eq.~\eqref{eq:QCD-Q}].
    Red curve: the NNLO result
    from in-medium chiral perturbation~\cite{
      Goda2013Phys.Rev.C88_065204}.
    The central values of $ C_1 $ and $ \sigma_{\pi N} $
    are taken for these curves.}
  \label{fig:QCD-mass}
\end{figure}

We decompose the mass difference between mirror nuclei
$ \Delta E = E \left(  Z + 1, N \right) - E \left( Z, N + 1 \right) $
into the Coulomb HF contribution $ \Delta E_{\urm{C}} $ and the ONS anomaly $ \delta_{\urm{ONS}} $ as 
\begin{equation}
  \label{eq:eq11}
  \Delta E
  =
  \Delta E_{\urm{C}}
  +
  \delta_{\urm{ONS}}. 
\end{equation}
On the basis of Eq.~\eqref{eq:mass-pn},
the CSB effect to $ \delta_{\urm{ONS}} $
from the partial restoration of chiral symmetry in the uniform and symmetric
($ N = Z $)
nuclear matter $ \delta_{\urm{chiral}} $ can be estimated as~\cite{
  Hatsuda1991Phys.Rev.Lett.66_2851}
\begin{equation}
  \label{eq:eq9}
  \delta_{\urm{chiral}}
  \equiv
  \Delta_{np} \left( 0 \right)
  -
  \Delta_{np} \left( \rho \right)
  =
  C_1
  \left[
    1
    - 
    G \left( \rho \right)
  \right].
\end{equation}
Shown in Fig.~\ref{fig:QCD-mass} is $ \delta_{\urm{chiral}} $
as a function of the baryon density by taking the central values of
$ C_1 $ and $ \sigma_{\pi N} $ mentioned above. 
Two curves correspond to the LO result with the Fermi-motion correction (LO*) in Eq.~\eqref{eq:QCD-Q}
and the next-to-next-to-leading order (NNLO) result
from the in-medium chiral perturbation~\cite{
  Goda2013Phys.Rev.C88_065204}.
The figure shows that Eq.~\eqref{eq:QCD-Q} is quite accurate at least up to $ \rho/\rho_0 \lesssim 1 $.
It should be noted here that $ \delta_{\urm{chiral}} $ is around a few hundreds $ \mathrm{keV} $
for $ \rho < \rho_0 $,
which is the right sign and magnitude to explain the ONS anomaly in finite nuclei.
\par
Let us make an alternative evaluation of $ \delta_{\urm{ONS}} $ in Eq.~\eqref{eq:eq11} on the basis of the CSB interaction of an energy density functional (EDF).
First of all, the general form of $ E \left( Z, N \right) $
for uniform nuclear matter up to the second order of $ \beta = \left( N - Z \right)/A$ reads~\cite{
  Naito:2023fnm}
\begin{equation}
  \label{eq:total-energy}
  \frac{E}{A}
  \simeq
  \varepsilon_0 \left( \rho \right)
  +
  \varepsilon_1 \left( \rho \right)
  \beta
  +
  \varepsilon_2 \left( \rho \right)
  \beta^2 .
\end{equation}
In particular, for $ N = Z $, 
we find the 
$ \left. \Delta E \right|_{N=Z} = - 2 \varepsilon_1 \left( \rho \right) $ 
where the effect of $ \varepsilon_0 $ and $ \varepsilon_2 $ disappears.
Note that $ \varepsilon_1 $ is a genuinely CSB-type term coming only from the CSB EDF.
In this Letter, we take the Skyrme-type CSB interaction~\cite{
  Sagawa2019Eur.Phys.J.A55_227}
to evaluate its contribution to $ \varepsilon_1 \left( \rho \right) $:
\begin{align}
  V_{\urm{CSB}} \left( \ve{r} \right)
  & =
    \left[
    \vphantom{
    \frac{s_1}{2}
    \left( 1 + y_1 P_{\sigma} \right)
    \left(
    \ve{k}^{\dagger 2} 
    \delta \left( \ve{r} \right)
    +
    \delta \left( \ve{r} \right)
    \ve{k}^2
    \right)}
    s_0 
    \left( 1 + y_0 P_{\sigma} \right)
    \delta \left( \ve{r} \right)
    \right.
    \notag \\
  & \quad
    \left.
    +
    \frac{s_1}{2}
    \left( 1 + y_1 P_{\sigma} \right)
    \left(
    \ve{k}^{\dagger 2} 
    \delta \left( \ve{r} \right)
    +
    \delta \left( \ve{r} \right)
    \ve{k}^2
    \right)
    \right.
    \notag \\
  & \quad
    \left.
    \vphantom{
    \frac{s_1}{2}
    \left( 1 + y_1 P_{\sigma} \right)
    \left(
    \ve{k}^{\dagger 2} 
    \delta \left( \ve{r} \right)
    +
    \delta \left( \ve{r} \right)
    \ve{k}^2
    \right)}
    +
    s_2
    \left( 1 + y_2 P_{\sigma} \right)
    \ve{k}^{\dagger}
    \cdot
    \delta \left( \ve{r} \right)
    \ve{k}
    \right]
    \frac{\tau_{1z} + \tau_{2z}}{4},   
    \label{eq:int_CSB}
\end{align}
where
$ \tau_{iz} = +1 $ ($ -1 $) for neutrons (protons) is
the $ z $ direction of isospin operator of nucleon $ i $,
$ \ve{k} = \left( \ve{\nabla}_1 - \ve{\nabla}_2 \right) / 2i $,
$ \ve{r} = \ve{r}_1 - \ve{r}_2 $, 
and $ P_{\sigma} = \left( 1 + \ve{\sigma}_1 \cdot \ve{\sigma}_2 \right) / 2 $
is the spin-exchange operator.
In Eq.~\eqref{eq:int_CSB},
$ s_0 $ and $ y_0 $ are the strength parameters of the contact CSB and its spin exchange interactions,
while $ s_1 $ ($ s_2 $) and $ y_1 $ ($ y_2 $) are the parameters of the momentum dependent $ s $-wave ($ p $-wave) CSB and its spin exchange interactions, respectively.  
Equation~\eqref{eq:int_CSB} gives contributions to
$ \varepsilon_1 \left( \rho \right) $ and hence $ \delta_{\urm{ONS}} $ as~\cite{
  Naito:2023fnm}
\begin{equation}
  \label{eq:SE-CSB}
  \delta_{\urm{Skyrme}}
  =
  -
  \frac{\tilde{s}_0}{4}
  \rho
  -
  \frac{1}{10}
  \left( \frac{3 \pi^2}{2} \right)^{2/3}
  \left(
    \tilde{s}_1
    +
    3 \tilde{s}_2
  \right)
  \rho^{5/3},
\end{equation}
where we have defined the effective coupling strengths,
\begin{equation}
  \tilde{s}_0
  \equiv
  s_0 \left( 1 - y_0 \right),
  \quad
  \tilde{s}_1
  \equiv
  s_1 \left( 1 - y_1 \right),
  \quad
  \tilde{s}_2
  \equiv
  s_2 \left( 1 + y_2 \right).
\end{equation}
Note that the Thomas-Fermi approximation is adopted to evaluate the kinetic energy terms in Eq.~\eqref{eq:total-energy}.
\par
There have been attempts to extract $ \tilde{s}_{\urm{$ 0 $, $ 1 $, $ 2 $}} $ by using various experimental data such as the energy of isobaric analog states (IAS)~\cite{
  Roca-Maza2018Phys.Rev.Lett.120_202501}
and the mass differences of mirror and isotriplet nuclei~\cite{
  Baczyk2019J.Phys.G46_03LT01}.
The value of $ \tilde{s}_0 $ estimated from IAS in $ \nuc{Pb}{208}{} $
is  $ \tilde{s}_0 = -52.6 \pm 1.4 \, \mathrm{MeV} \, \mathrm{fm}^3 $,
while the mass differences of mirror nuclei lead to two estimates;
$ \left( \tilde{s}_0, \tilde{s}_{\urm{$ 1 $, $ 2 $}} \right) =
\left( -29.2 \pm 1.2 \, \mathrm{MeV} \, \mathrm{fm}^3, 0 \right) $
and
$ \left( \tilde{s}_0, \tilde{s}_1, \tilde{s}_2 \right) =
\left( 44 \pm 8 \, \mathrm{MeV} \, \mathrm{fm}^3, -56 \pm 16 \, \mathrm{MeV} \, \mathrm{fm}^5, -31.2 \pm 3.2 \, \mathrm{MeV} \, \mathrm{fm}^5 \right) $.
The parameters in Ref.~\cite{
  Baczyk2019J.Phys.G46_03LT01}
are related to ours as $ t^{\urm{III}}_{\urm{$ 0 $, $ 1 $, $ 2 $}} = \tilde{s}_{\urm{$ 0 $, $ 1 $, $ 2 $}} /4 $.
Since the contributions of 
$ \tilde{s}_0 $ and $ \tilde{s}_{\urm{$ 1 $, $ 2 $}} $ tend to cancel each other
in physical observables,
it is rather difficult to determine 
the magnitude and the sign
of each term only from the present experimental data.
\par
On the other hand, our approach is to constrain
$ \tilde{s}_{\urm{$ 0 $, $ 1 $, $ 2 $}} $
from the low-energy constants in QCD, $ \gamma $ and $ \sigma_{\pi N} $,
by matching $ \delta_{\urm{Skyrme}} \left( \rho \right) $ in Eq.~\eqref{eq:SE-CSB}
and $ \delta_{\urm{chiral}} \left( \rho \right) $
expanded up to $ \mathcal{O} \left( \rho^{5/3} \right) $ at low densities.
Then, we obtain
\begin{equation}
  \tilde{s}_0
  =
  - \frac{4}{3} 
  \frac{C_1 \sigma_{\pi N}}{f_{\pi}^2 m_{\pi}^2} ,
  \qquad
  \tilde{s}_1 + 3 \tilde{s}_2
  =
  \frac{1}{m_N^2}\frac{C_1 \sigma_{\pi N}}{f_{\pi}^2 m_{\pi}^2}.
\end{equation}
The magnitudes and signs of $ \tilde{s}_0 $ and $ \tilde{s}_1 + 3 \tilde{s}_2 $ are
summarized in Table \ref{tab:tab-csb},
where the linear uncertainty estimation is used.
To evaluate the CSB effect in finite nuclei,
where $ \tilde{s}_1 $ and $ \tilde{s}_2 $ contribute independently, two characteristic parameter sets (Cases I and II) are introduced.
\begin{table}[b]
  \centering
  \caption{
    Parameters of the Skyrme-type CSB interactions
    constrained from the low-energy constants in QCD.
    To evaluate the CSB effect in finite nuclei, where 
    $ \tilde{s}_1 $ and $ \tilde{s}_2 $ contribute independently,
    two characteristic parameter sets (Cases I and II) are introduced.}
  \label{tab:tab-csb}
  \begin{ruledtabular}
    \begin{tabular}{ldd}
      $ \tilde{s}_0 $ ($ \mathrm{MeV} \, \mathrm{fm}^3 $)
      & -15.5^{+8.8}_{-12.5} & \\   
      $ \tilde{s}_1 + 3 \tilde{s}_2 $  ($ \mathrm{MeV} \, \mathrm{fm}^5 $)
      & 0.52^{+0.42}_{-0.29} & \\
      \hline
      & \multicolumn{1}{c}{Case I} & \multicolumn{1}{c}{Case II} \\ \hline
      $ \tilde{s}_0 $ ($ \mathrm{MeV} \, \mathrm{fm}^3 $) & -15.5^{+8.8}_{-12.5} & -15.5^{+8.8}_{-12.5} \\
      $ \tilde{s}_1 $ ($ \mathrm{MeV} \, \mathrm{fm}^5 $) & 0.52^{+0.42}_{-0.29} & 0.00 \\  
      $ \tilde{s}_2 $ ($ \mathrm{MeV} \, \mathrm{fm}^5 $) & 0.00                 & 0.18^{+0.14}_{-0.10} \\
    \end{tabular}
  \end{ruledtabular}
\end{table}
\begin{table}[t]
  \centering
  \caption{
    The neutron radiii, the proton ones, and the charge ones of
    $ \nuc{O}{16}{} $ and $ \nuc{Ca}{40}{} $.
    Two types of Skyrme EDFs SGII and SAMi,
    are adopted for the HF calculation.
    Experimental data are taken from Refs.~\cite{
      DeVries1987At.DataNucl.DataTables36_495,
      Zenihiro:2018rmz,Fricke2004}.}
  \label{tab:tab-radius}
  \begin{ruledtabular}
    \begin{tabular}{lddd}
      \multicolumn{1}{c}{$ \nuc{O}{16}{} $} & \multicolumn{1}{c}{$ r_n $} & \multicolumn{1}{c}{$ r_p $} & \multicolumn{1}{c}{$ r_c $} \\
      \hline
      SGII & 2.601 & 2.626 & 2.744 \\
      SAMi & 2.625 & 2.648 & 2.765 \\
      Expt.~\cite{DeVries1987At.DataNucl.DataTables36_495} & \multicolumn{1}{c}{---} & \multicolumn{1}{c}{---} & 2.737 \\
      \hline
      \multicolumn{1}{c}{$ \nuc{Ca}{40}{} $} & \multicolumn{1}{c}{$ r_n $} & \multicolumn{1}{c}{$ r_p $} & \multicolumn{1}{c}{$ r_c $} \\
      \hline
      SGII & 3.325 & 3.374 & 3.467 \\
      SAMi & 3.342 & 3.390 & 3.482 \\
      Expt.~\cite{Zenihiro:2018rmz} & 3.375 & 3.385 & 3.480 \\
      Expt.~\cite{Fricke2004} & \multicolumn{1}{c}{---} & \multicolumn{1}{c}{---} & \multicolumn{1}{c}{$3.478$} \\
    \end{tabular}
  \end{ruledtabular}
\end{table}
\par
To carry out precise calculation of the mass differences of mirror nuclei,
we consider two types of the Skyrme EDFs
for the isospin symmetric part,
SGII~\cite{
  VanGiai1981Phys.Lett.B106_379}
and SAMi~\cite{
  Roca-Maza2012Phys.Rev.C86_031306};
they reproduce well, within 0.3\%, the experimental radii of the
$ N = Z $ closed shell nuclei,
$ \nuc{O}{16}{} $ and $ \nuc{Ca}{40}{} $,
as shown in Table~\ref{tab:tab-radius}.
It is important for any adopted EDF to reproduce the charge radii 
since the Coulomb energy part $ \Delta E_{\urm{C}} $
is essentially determined by the charge distribution:
The change of $ 1 \, \% $ in the charge radius of $ \nuc{Ca}{40}{} $
gives rise to $ 20 $--$ 30 \, \mathrm{keV} $
difference in $ \Delta E_{\urm{C}} $ of mirror nuclei.   
\begin{table}[t]
  \centering
  \caption{
    Contributions from the Skyrme CSB interactions to $ \delta_{\urm{ONS}} $
    in Cases I and II with theoretical uncertainties.
    The values are given in unit of $ \mathrm{keV} $.
    The core density and the wave function of valence orbit are calculated by HF model with Skyrme EDFs, SGII and SAMi.
    All the values are obtained self-consistently.}
  \begin{ruledtabular}
    \label{tab:tab0}
    \begin{tabular}{l|lllll}
      &
        \multicolumn{1}{l}{Nuclei} & \multicolumn{1}{c}{$ \nuc{F}{17}{} $-$ \nuc{O}{17}{} $} & \multicolumn{1}{c}{$ \nuc{O}{15}{} $-$ \nuc{N}{15}{} $} & \multicolumn{1}{c}{$ \nuc{Sc}{41}{} $-$ \nuc{Ca}{41}{} $} & \multicolumn{1}{c}{$ \nuc{Ca}{39}{} $-$ \nuc{K}{39}{} $} \\
      \hline
      & Orbital & \multicolumn{1}{c}{$ 1d_{5/2} $} & \multicolumn{1}{c}{$ \left( 1p_{1/2} \right)^{-1} $} & \multicolumn{1}{c}{$ 1f_{7/2} $} & \multicolumn{1}{c}{$ \left( 1d_{3/2} \right)^{-1} $} \\
      \hline
      \multirow{5}{*}{SGII}
      & $ \tilde{s}_0 $ & $ 229^{+192}_{-125} $ & $ 269^{+221}_{-148} $ & $ 292^{+245}_{-160} $  & $ 322^{+264}_{-176} $ \\
      & $ \tilde{s}_1 $ ($ \tilde{s}_2 = 0 $) & $ -5.0^{+2.8}_{-4.0} $ & $ -5.6^{+3.1}_{-4.5} $ & $ -6.6^{+3.7}_{-5.3} $ & $ -6.0^{+3.4}_{-4.9} $ \\
      & $ \tilde{s}_2 $ ($ \tilde{s}_1 = 0 $) & $ -6.4^{+3.5}_{-5.2} $ & $-3.3^{+1.8}_{-2.7}$ & $-5.3^{+2.9}_{-4.3}$ & $-5.0^{+2.8}_{-4.1}$ \\
      & Case I  & $ 224^{+192}_{-125} $ & $ 264^{+221}_{-148} $ & $ 287^{+245}_{-160} $ & $ 315^{+264}_{-176} $ \\
      & Case II & $ 225^{+192}_{-125} $ & $ 266^{+221}_{-148} $ & $ 289^{+245}_{-160} $ & $ 316^{+264}_{-176} $ \\
      \hline
      \multirow{5}{*}{SAMi}
      & $ \tilde{s}_0 $ & $ 211^{+174}_{-115} $ & $ 274^{+225}_{-152} $ & $ 278^{+230}_{-151} $ & $ 324^{+269}_{-180} $ \\
      & $ \tilde{s}_1 $ ($ \tilde{s}_2 = 0 $) & $ -5.2^{+2.9}_{-4.2} $ & $ -5.4^{+3.0}_{-4.4} $ & $ -7.3^{+4.0}_{-5.9} $ & $ -8.4^{+4.6}_{-6.6} $ \\
      & $ \tilde{s}_2 $ ($ \tilde{s}_1 = 0 $) & $ -4.1^{+2.3}_{-3.3} $ & $ -3.2^{+1.8}_{-2.6} $ & $ -5.7^{+3.1}_{-4.6} $ & $ -5.2^{+2.9}_{-4.2} $ \\
      & Case I  & $ 206^{+174}_{-115} $ & $ 269^{+225}_{-152} $ & $ 271^{+230}_{-151} $ & $ 321^{+269}_{-180} $ \\
      & Case II & $ 207^{+174}_{-115} $ & $ 271^{+225}_{-152} $ & $ 272^{+230}_{-151} $ & $ 322^{+269}_{-180} $ \\
    \end{tabular}
  \end{ruledtabular}
\end{table}
\par
The contributions of the CSB interactions to the mass difference  between
mirror nuclei of $ \left( N \pm 1, Z \right) $ and $ \left( N, Z \pm 1 \right) $ with the closed-shell core $ A = N + Z = 16 $ and $ 40 $ 
are calculated by using HF wave functions for SGII and SAMi.
As we can see from the Table~\ref{tab:tab0},
$ \tilde{s}_0 $ provides a dominant contribution ($ 210 $--$ 320 \, \mathrm{keV} $),
more than one order of magnitude larger than the $ \tilde{s}_{\urm{$ 1 $, $ 2 $}} $ contributions.
The net results are slightly different from the sum of
$ \tilde{s}_0 $ and $ \tilde{s}_{\urm{$ 1 $ ($ 2 $)}} $ contributions due to the nonlinear effect in the calculation using EDFs.
The final results of Case I and those of Case II 
are essentially identical due to the $ \tilde{s}_0 $ 
dominance, so that we focus on Case I below.
\par
Let us now turn to the comparison of the theoretical values with our CSB interaction
with the experimental mass difference of the mirror nuclei by including $ \Delta E_{\urm{C}} $ 
and other extra contributions~\cite{
  Uehling1935Phys.Rev.48_55,
  VanGiai1971Phys.Lett.B35_135,
  Suzuki1992Nucl.Phys.A536_141,
  Roca-Maza2018Phys.Rev.Lett.120_202501}
(see Supplemental Material~\cite{SM}).
The results are summarized in Table~\ref{tab:tab4} for SGII and Table~\ref{tab:tab5} for SAMi assuming Case I.
First, we note that the core density and the wave function of valence orbital are calculated with the closed shell core configuration without the core polarization effect of the valence nucleon.
The direct and exchange contributions of the Coulomb interaction
($ \Delta E_{\urm{D}} $ and $ \Delta E_{\urm{E}} $ with
$ \Delta E_{\urm{C}} = \Delta E_{\urm{D}} + \Delta E_{\urm{E}} $)
are obtained with the exact treatment of the exchange term.  
The sum of extra contributions including
the finite-size effect of nucleon,
the center-of-mass effect on nuclear density,
the Thomas-Ehrman effect $ \delta_{NN}^1 $,
the isospin impurity $ \delta_{NN}^2 $,
the electromagnetic spin-orbit interaction,
the core polarization effect of the last nucleon,
the proton and neutron mass difference in the kinetic energy,
and the vacuum polarization, are listed as ``Extra'' in the Tables~\ref{tab:tab4} and \ref{tab:tab5}: Each contribution varies  from $ -150 \, \mathrm{keV} $ to $ 150 \, \mathrm{keV} $,
while the net result is at most $ 100 \, \mathrm{keV} $ due to a strong cancellation.
See Supplemental Material~\cite{SM} for the details.
\par
\begin{figure}[tb]
  \centering
  \includegraphics[width=1.0\linewidth]{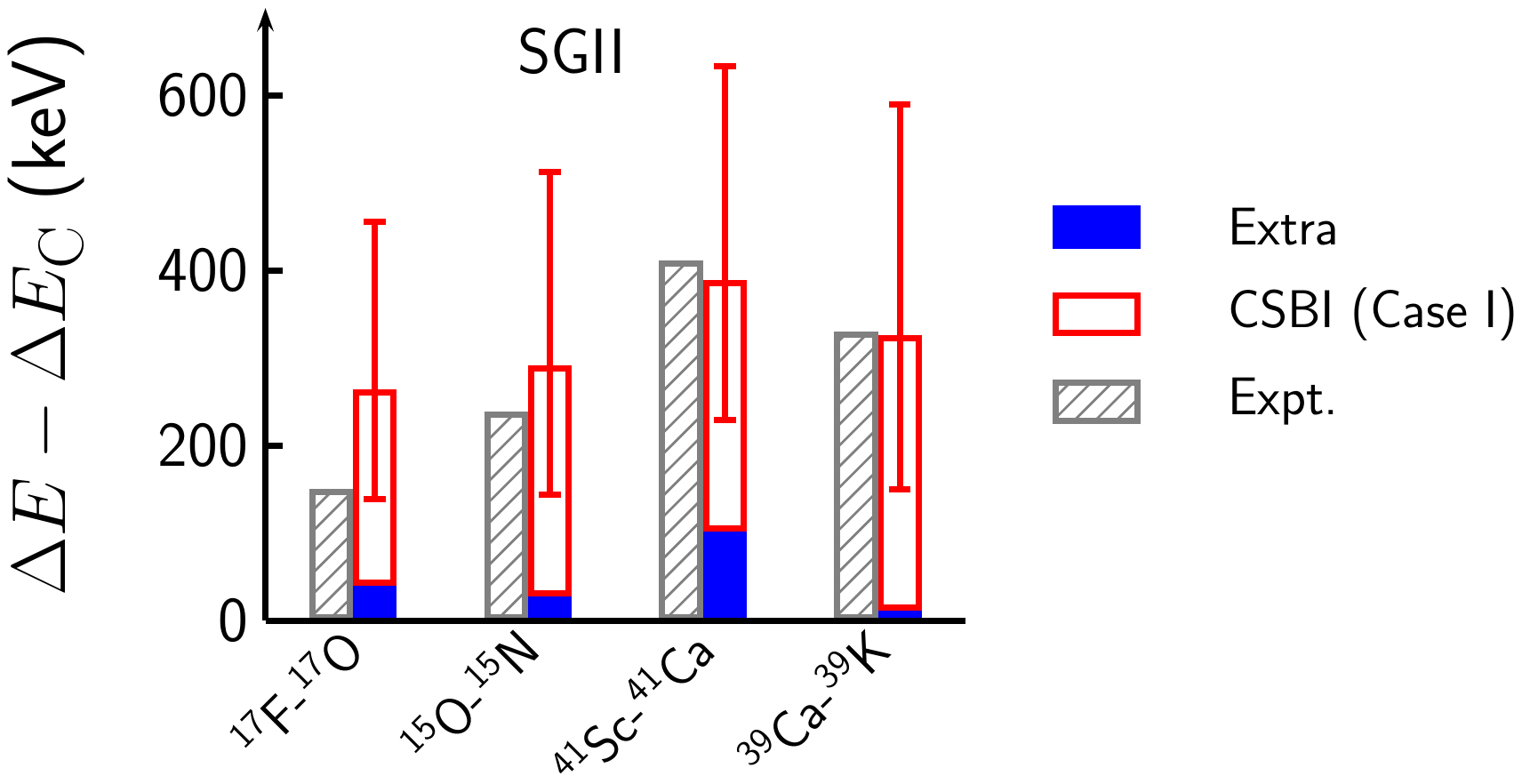}
  \includegraphics[width=1.0\linewidth]{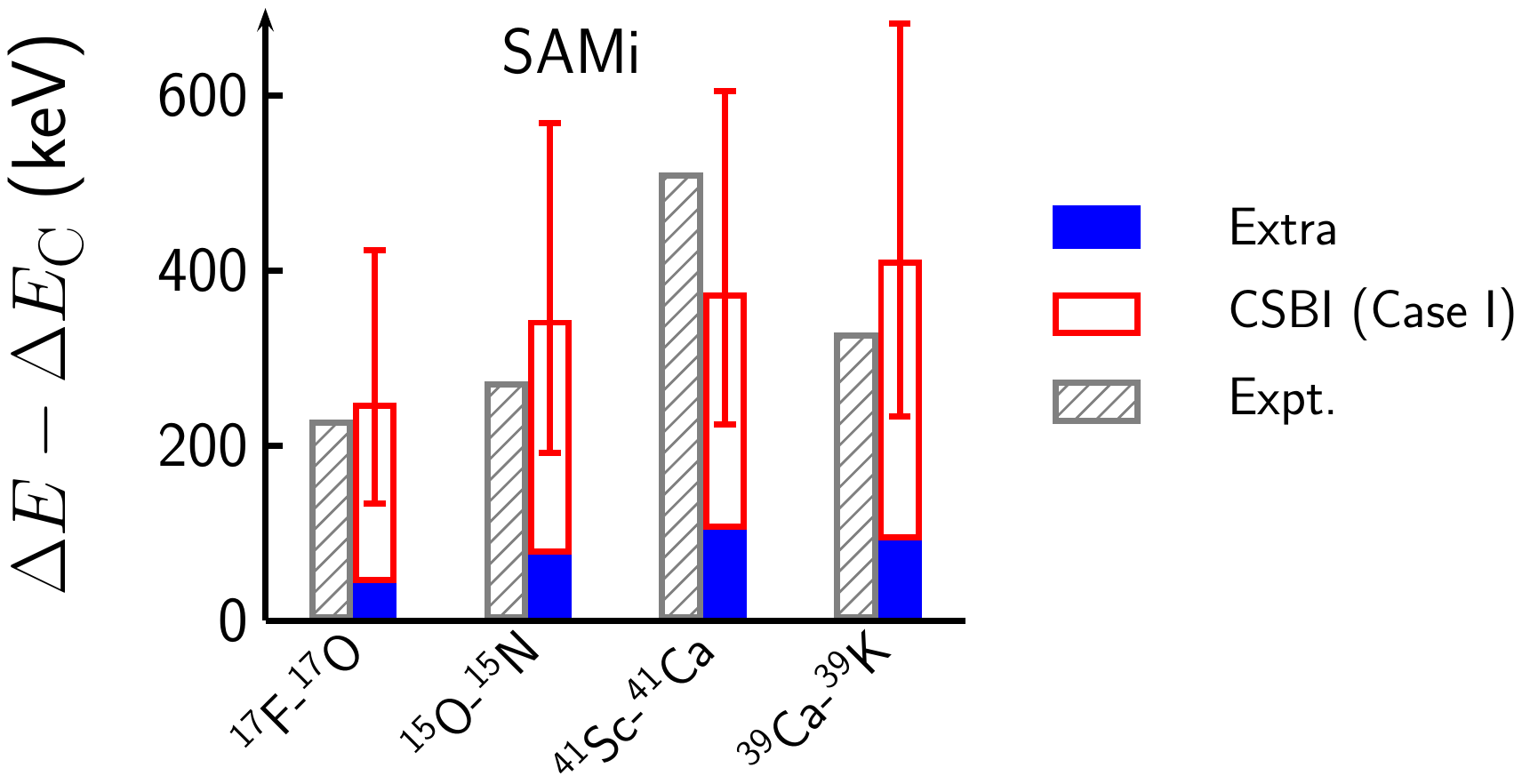}
  \caption{
    Comparisons of the experimental ONS anomaly
    $ \Delta E_{\urm{Expt.}} -\Delta E_{\urm{C}} $ (grey hatched bars)
    and the corresponding theoretical estimates in two EDFs (SGII and SAMi).
    The contribution from the QCD-based CSB interaction (CSBI) in Case I 
    and the extra contributions are
    indicated by the red bars with error bars and the blue bars, respectively.}
  \label{fig:anomaly-crop}
\end{figure}

The sum of $ \Delta E_{\urm{D}} $, $ \Delta E_{\urm{E}} $
and Extra denoted by ``Sum (without CSBI)'' in the tables is systematically 
smaller than the experimental value by $ 3 $--$ 6 \, \% $. 
Our CSBI contributions constrained by the low-energy constants in QCD 
fill the gap as it can be seen by comparing the sum neglecting CSBI effects,
the sum containing CSBI effects ``Sum (with CSBI)'' and the experimental (``Expt.'') rows in 
Tables~\ref{tab:tab4} and \ref{tab:tab5}. 
Shown in Fig.~\ref{fig:anomaly-crop} is
$ \delta_{\urm{ONS}} = \Delta E - \Delta E_{\urm{C}} $
where it is evident the agreement between experiment and the present theoretical estimates.  
The theoretical error bars given in Table \ref{tab:tab0}
are shown in the figure,
while the experimental error bars are around only $ 5 \, \mathrm{keV} $ and would not be visible in the scale of the figure. 
\begin{table}[t]
  \centering
  \caption{
    The breakdown of the  mass differences of mirror nuclei $ \Delta E $
    into each contribution Coulomb, Extra and CSB interaction (CSBI) for Case I with the Skyrme EDF SGII.
    Numbers are given in units of $ \mathrm{MeV} $.}
  \label{tab:tab4}
  \begin{ruledtabular}
    \begin{tabular}{ldddd}
      \multicolumn{1}{l}{Nuclei} & \multicolumn{1}{c}{$ \nuc{F}{17}{} $-$ \nuc{O}{17}{} $} & \multicolumn{1}{c}{$ \nuc{O}{15}{} $-$ \nuc{N}{15}{} $} & \multicolumn{1}{c}{$ \nuc{Sc}{41}{} $-$ \nuc{Ca}{41}{} $} & \multicolumn{1}{c}{$ \nuc{Ca}{39}{} $-$ \nuc{K}{39}{} $} \\
      \hline
      \multicolumn{1}{l}{Orbital} & \multicolumn{1}{c}{$ 1d_{5/2} $} & \multicolumn{1}{c}{$ \left( 1p_{1/2} \right)^{-1} $} & \multicolumn{1}{c}{$ 1f_{7/2} $} & \multicolumn{1}{c}{$ \left( 1d_{3/2} \right)^{-1} $} \\
      \hline
      $ \Delta E_{\urm{D}} $ (Coulomb) &  3.596 & 3.272 &  7.133 & 6.717 \\
      $ \Delta E_{\urm{E}} $ (Coulomb) & -0.203 & 0.026 & -0.267 & 0.260 \\
      Extra                            &  0.040 & 0.028 &  0.102 & 0.011 \\
      CSBI (Case I)                    &  0.224 & 0.264 &  0.287 & 0.315 \\
      \hline
      Sum (without CSBI)               &  3.432 & 3.326 &  6.965 & 6.985 \\ 
      Sum (with CSBI)                  &  3.656 & 3.590 &  7.252 & 7.300 \\
      \hline
      Expt.~\cite{Huang2021Chin.Phys.C45_030002} &  3.543 & 3.537 &  7.278 & 7.307 \\
    \end{tabular}
  \end{ruledtabular}
\end{table}
\begin{table}[t]
  \centering
  \caption{
    The same as Table \ref{tab:tab4}, but with the Skyrme EDF SAMi.}
  \label{tab:tab5}
  \begin{ruledtabular}
    \begin{tabular}{ldddd}
      \multicolumn{1}{l}{Nuclei} & \multicolumn{1}{c}{$ \nuc{F}{17}{} $-$ \nuc{O}{17}{} $} & \multicolumn{1}{c}{$ \nuc{O}{15}{} $-$ \nuc{N}{15}{} $} & \multicolumn{1}{c}{$ \nuc{Sc}{41}{} $-$ \nuc{Ca}{41}{} $} & \multicolumn{1}{c}{$ \nuc{Ca}{39}{} $-$ \nuc{K}{39}{} $} \\
      \hline
      \multicolumn{1}{l}{Orbital} & \multicolumn{1}{c}{$ 1d_{5/2} $} & \multicolumn{1}{c}{$ \left( 1p_{1/2} \right)^{-1} $} & \multicolumn{1}{c}{$ 1f_{7/2} $} & \multicolumn{1}{c}{$ \left( 1d_{3/2} \right)^{-1} $} \\
      \hline
      $ \Delta E_{\urm{D}} $ (Coulomb) &  3.506 & 3.242 &  7.025 & 6.697 \\
      $ \Delta E_{\urm{E}} $ (Coulomb) & -0.193 & 0.022 & -0.259 & 0.281 \\
      Extra                            &  0.043 & 0.075 &  0.104 & 0.092 \\ 
      CSBI (Case I)                    &  0.206 & 0.269 &  0.271 & 0.321 \\
      \hline
      Sum (without CSBI)               &  3.356 & 3.339 &  6.870 & 7.070 \\
      Sum (with CSBI)                  &  3.562 & 3.608 &  7.141 & 7.391 \\
      \hline
      Expt.~\cite{Huang2021Chin.Phys.C45_030002} &  3.543 & 3.537 &  7.278 & 7.307 \\
    \end{tabular}
  \end{ruledtabular}
\end{table}
\par
Finally, as mentioned, there has been a recent effort in quantifying the effects of CSB in some selected nuclear observables~\cite{
  Roca-Maza2018Phys.Rev.Lett.120_202501,
  Baczyk2018Phys.Lett.B778_178,
  Baczyk2019J.Phys.G46_03LT01,
  Sagawa2022Phys.Lett.B829_137072,
  Novario2023Phys.Rev.Lett.130_032501}.
However, depending on the theoretical method employed, the estimated central values of the leading order term CSB parameter ($ \tilde{s}_0 $) can differ by one order of magnitude among the different approaches and could even be of different sign (cf.~Ref.~\cite{
  Naito2023}
and Supplemental Material~\cite{SM}).
\par
In summary, we evaluated the EDF parameters of Skyrme-type CSB interactions,
not only the contact term ($ \tilde{s}_0 $)
but also the momentum-dependent terms ($ \tilde{s}_{\urm{$ 1 $ ($ 2 $)}} $), 
by utilizing the low-energy constants in QCD and the density dependence of chiral condensation of $\bar{q}q$ pair in the nuclear medium for the first time.
The resulting QCD-based CSB interaction is applied to resolve the ONS anomaly:
The numerical results for the mirror nuclei
($ A = 16 \pm 1 $ and $ A = 40 \pm 1 $ with the isosymmetric core $ N = Z = A/2 $)
with two Skyrme EDFs (SGII and SAMi)
show good agreement with experimental data both in sign and magnitude within the theoretical 
error bars.
Major theoretical uncertainty of the final results 
originates from the values of $ \gamma $ and $ \sigma_{\pi N} $:
Increasing the accuracy of these constants from the experimental data or from the lattice QCD simulations will be instrumental.
\par
The QCD-based CSB interaction discussed in this Letter would have strong impact on isospin symmetry breaking phenomena such as IAS, the super-allowed $ \beta $ decay in the context of Cabibbo-Kobayashi-Maskawa unitary matrix, and the mass predictions of mirror and isotriplet nuclei near the proton drip line.
We plan to make systematic studies of these quantities. 
\begin{acknowledgments}
  The authors thank Gianluca Col\`o and Toshio Suzuki for fruitful discussions.
  H.~S.~acknowledges the Grant-in-Aid for Scientific Research (C) under Grant No.~19K03858.
  T.~N.~acknowledges
  the RIKEN Special Postdoctoral Researcher Program,
  the JSPS Grant-in-Aid for Research Activity Start-up under Grant No.~22K20372,
  the JSPS Grant-in-Aid for Transformative Research Areas (A) under Grant No.~23H04526,
  the JSPS Grant-in-Aid for Scientific Research (B) under Grant No.~23H01845,
  and
  the JSPS Grant-in-Aid for Scientific Research (C) under Grant No.~23K03426.
  T.~H.~was partially supported by JSPS Grant-in-Aid No.~18H05236.
\end{acknowledgments}
\end{document}

% 
%%%%%%%%%%%%%%%%%%%%%%%%%%%%%%%%%%%%%%%%%%%%%%%%%% 
\begin{CJK*}{UTF8}{}
  \title{Supplemental Material: \\
    QCD-based charge symmetry breaking interaction \\
    and Okamoto-Nolen-Schiffer anomaly}
  \author{Hiroyuki Sagawa (\CJKfamily{min}{佐川弘幸})}
  \affiliation{
    RIKEN Nishina Center for Accelerator-based Science, Wako 351-0198, Japan}
  \affiliation{
    Center for Mathematics and Physics, University of Aizu, 
    Aizu-Wakamatsu, Fukushima 965-8560, Japan}
  \author{Tomoya Naito (\CJKfamily{min}{内藤智也})}
  \affiliation{
    RIKEN Interdisciplinary Theoretical and Mathematical Sciences Program (iTHEMS),
    Wako 351-0198, Japan}
  \affiliation{
    Department of Physics, Graduate School of Science, The University of Tokyo,
    Tokyo 113-0033, Japan}
  \author{Xavier Roca-Maza}
  \affiliation{
    Departament de F\'{\i}sica Qu\`{a}ntica i Astrof\'{\i}sica
    and
    Institut de Ci\`{e}ncies del Cosmos,
    Universitat de Barcelona, 
    Mart\'{\i} i Franqu\'{e}s 1,
    08028 Barcelona, Spain}
  \affiliation{
    Dipartimento di Fisica, Universit\`{a} degli Studi di Milano,
    Via Celoria 16, 20133 Milano, Italy}
  \affiliation{
    INFN, Sezione di Milano,
    Via Celoria 16, 20133 Milano, Italy}
  \author{Tetsuo Hatsuda (\CJKfamily{min}{初田哲男})}
  \affiliation{
    RIKEN Interdisciplinary Theoretical and Mathematical Sciences Program (iTHEMS),
    Wako 351-0198, Japan}
  \date{\today}
  %%%%%%%%%%%%%%%%%%%%%%%%%%%%%%%%%%%%%%%%%%%%%%%%%% 
  \begin{abstract}
    In this supplemental material,
    the fitting procedure of the chiral-condensate as well as 
    the numerical details of various contributions
    to the mass differences of mirror nuclei in the Hartree-Fock calculation~\cite{
      VanGiai1971Phys.Lett.B35_135,
      Suzuki1992Nucl.Phys.A536_141}
    are presented.    
  \end{abstract}
  \maketitle
\end{CJK*}
%%%%%%%%%%%%%%%%%%%%%%%%%%%%%%%%%%%%%%%%%%%%%%%%%% 
\section{In-medium chiral condensate}
\par
\begin{figure}[tb]
  \centering
  \includegraphics[width=0.5\linewidth]{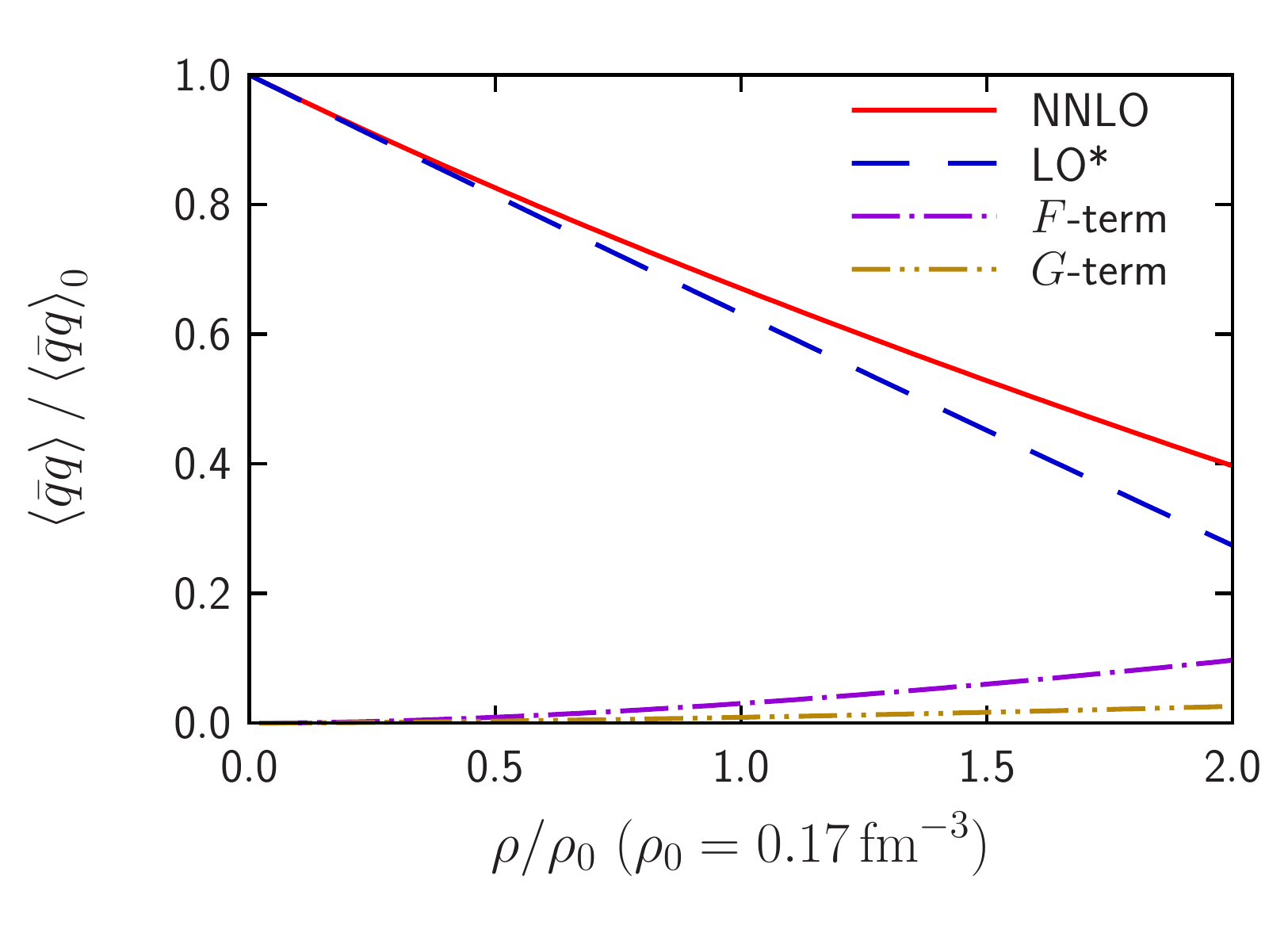}
  \caption{
    The in-medium chiral condensate $ \avr{\bar{q} q}/\avr{\bar{q} q}_0 $ obtained by Ref.~\cite{
      Goda2013Phys.Rev.C88_065204}.
    The red solid line corresponds to the full NNLO result in Eq.~\eqref{eq:QCD_0}, while
    blue long-dashed line corresponds to the result of LO* in Eq.~\eqref{eq:QCD_1}.
    The purple dot-dashed line and the  yellow dot-dot-dashed line correspond to the $ F $-term and the $ G $-term, respectively.}
  \label{fig:fitting_Goda}
\end{figure}

Figure~\ref{fig:fitting_Goda} shows the next-to-next-to-leading-order (NNLO) chiral condensate 
in nuclear matter obtained by the in-medium chiral perturbation theory~\cite{Goda2013Phys.Rev.C88_065204}; 
\begin{widetext}
  \begin{equation}
    \label{eq:QCD_0}
    \frac{\avr{\bar{q} q}}{\avr{\bar{q} q}_0}
    =
    1
    -
    \frac{\sigma_{\pi N}}{f_{\pi}^2 m_{\pi}^2}
    \rho
    \left( 1 - \frac{3 k_{\urm{F}}^2}{10 m_N^2} \right)
    +
    \frac{g_A^2 k_{\urm{F}}^4}{4 f_{\pi}^4 \pi^4}
    F \left( \frac{m_{\pi}^2}{4 k_{\urm{F}}^2} \right)
    +
    \frac{2 \sigma_{\pi N}^2 k_{\urm{F}}^4}{3 f_{\pi}^4 \pi^4 m_{\pi}^2}
    G \left( \frac{m_{\pi}^2}{4 k_{\urm{F}}^2} \right),
  \end{equation}
  with $ k_{\urm{F}} = \left( 3\pi^2 \rho/2 \right)^{1/3} $ and
  \begin{align}
    F \left( a^2 \right)
    & =
      \frac{3}{8}
      -
      \frac{3}{4} a^2
      -
      \frac{3a}{2}
      \arctan
      \frac{1}{a}
      +
      \frac{3}{4}
      a^2
      \left( a^2 + 2 \right)
      \log
      \frac{a^2 + 1}{a^2}, \\
    G \left( a^2 \right)
    & = 
      \frac{3}{8}
      -
      \frac{1}{4} a^2
      -
      a 
      \arctan
      \frac{1}{a}
      +
      \frac{1}{4}
      a^2
      \left( a^2 + 3 \right)
      \log
      \left| 
      \frac{1 + a^2}{a^2}
      \right|.
  \end{align}
\end{widetext}
Here, we adopt 
the pion mass $ m_{\pi} = 135 \, \mathrm{MeV} $,
the nucleon mass $ m_N = 938 \, \mathrm{MeV} $, 
the nuclear saturation density $ \rho_0 = 0.17 \, \mathrm{fm}^{-3} $,
the pion decay constant $ f_{\pi} = 92.4 \, \mathrm{MeV} $,
and the axial-vector coupling constant $ g_A = 1.26 $.
As seen in Fig.~\ref{fig:fitting_Goda},
the $ F $- and $ G $-terms originating from the two-loop order
are numerically small compared to the whole $ \avr{\bar{q} q} / \avr{\bar{q} q}_0 $ at the normal density $ \rho = \rho_0 $.
Hence, we neglect these contributions and consider only the leading-order term with the Fermi motion correction (called in LO* in the main text): 
\begin{equation}
  \label{eq:QCD_1}
  \frac{\avr{\bar{q} q}}{\avr{\bar{q} q}_0}
  \simeq
  1
  -
  \frac{\sigma_{\pi N}}{f_{\pi}^2 m_{\pi}^2}
  \rho
  \left( 1 - \frac{3 k_{\urm{F}}^2}{10 m_N^2} \right).
\end{equation}
\par
\begin{figure}[tb]
  \centering
  \includegraphics[width=0.5\linewidth]{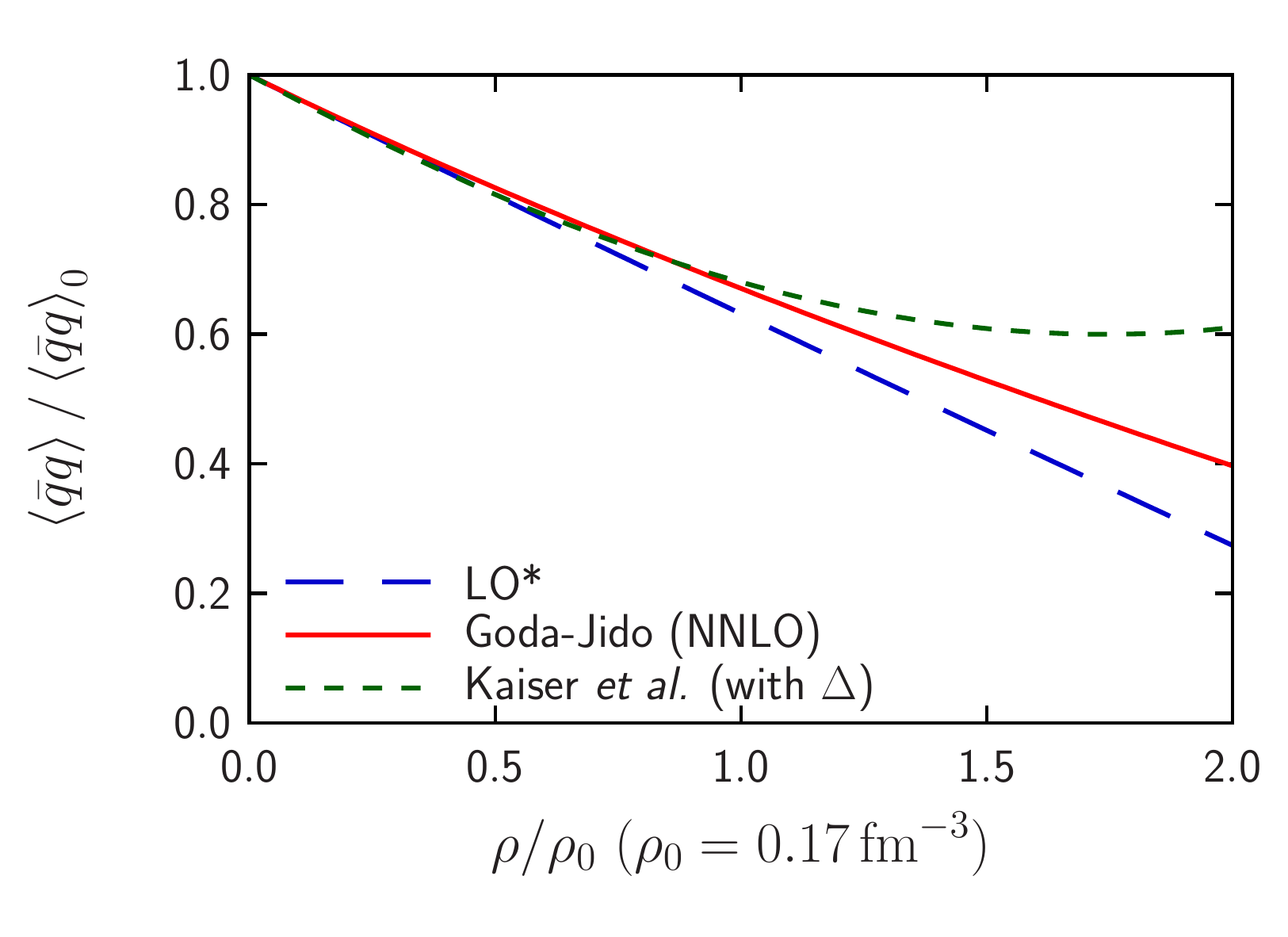}
  \caption{
    The red solid line corresponds to the full NNLO result in Eq.~\eqref{eq:QCD_0}, while
    green-dashed line corresponds to the
    result of Ref.~\cite{
      Kaiser2008Phys.Rev.C77_025204}
    with the $ \Delta $-excitation.
    The blue long-dashed line is the
    result of LO* in Eq.~\eqref{eq:QCD_1}.
    In all three cases,
    $ \sigma_{\pi N} = 45 \, \mathrm{MeV} $ is taken.}
  \label{fig:Kaiser}
\end{figure}

In order to check the model dependence on higher-order terms beyond LO*, we consider an alternative calculation of the in-medium chiral condensate, as detailed in Ref.~\cite{Kaiser2008Phys.Rev.C77_025204}.
Shown in Fig.~\ref{fig:Kaiser}
is a comparison of the LO*, analytic NNLO result in Eq.~\eqref{eq:QCD_0}~\cite{
  Goda2013Phys.Rev.C88_065204}
and the numerical result of Ref.~\cite{
  Kaiser2008Phys.Rev.C77_025204}
with the $ \Delta $-excitation. 
The comparison indicates good agreement among all three cases, especially below $ \rho_0 $.
This finding supports the validity of our LO* analysis in the main text.
\section{Coulomb contributions to $ \Delta E $ in the Hartree-Fock calculation}
\par
The direct term of the Coulomb interaction $ E_{\urm{D}} $ is calculated by using
the valence-proton wave function $ \psi_j^p \left( \ve{r} \right) $ of a $ j $-orbital as
\begin{widetext}
  \begin{align}
    E_{\urm{D}} \left( j \right)
    & =
      \int
      V_{\urm{C}} \left( \ve{r} \right)
      \left|
      \psi_j^p \left( \ve{r} \right)
      \right|^2
      \, d \ve{r}
      \notag \\
    & =
      \int d \ve{r}''
      \int d \ve{r}'''
      \,
      \frac{\rho_p \left( \ve{r}''' \right)}{\left| \ve{r}'' - \ve{r}''' \right|}
      \left| \psi_j^p \left( \ve{r}'' \right) \right|^2
      \sum_{j' \in \text{core}}
      \int_0^{\infty} r^2 \, dr
      \int_0^{\infty} r'^2 \, dr'
      \,
      R_j^2 \left( r \right)
      R_{j'}^2 \left( r' \right)
      \frac{1}{r_{>}},
      \label{eq:h1}
  \end{align}
  where $ R_j \left( r \right) $ is the radial part of the proton wave function
  and $ r_{>} $ is the larger of either $ r $ or $ r' $.
  The exact exchange term $ E_{\urm{E}} $ is given by
  \begin{equation}
    \label{eq:h2}
    E_{\urm{E}} \left( j \right)
    =
    - \sum_{j' \in \text{core}}
    \sum_k
    \left( 2j' + 1 \right)^{-1}
    \braket{j \frac{1}{2} k 0}{j' \frac{1}{2}}
    \int_0^{\infty} r^2 \, dr
    \int_0^{\infty} r'^2 \, dr'
    \,
    R_j \left( r \right)
    R_{j'} \left( r' \right)
    R_j \left( r' \right)
    R_{j'} \left( r \right)
    \frac{r_{<}^k}{r_{>}^{k+1}},
  \end{equation}
\end{widetext}
where $ \braket{j \frac{1}{2} k 0}{j' \frac{1}{2}} $ is the Clebsch-Gordan coefficient,
$ r_{<} $ is the smaller of either $ r $ or $ r' $,
and the selection rule, $ \left| l_j + k + l_{j'} \right| = \text{even} $,
should be hold.
Equation~\eqref{eq:h2} is exact in the case $ j \neq j' $,
i.e., for the particle state.
We take into account the effect of anti-symmetrization properly
for the case $ j = j' $,
which occurs in the calculations of hole states, i.e.,
we should take care of a factor $ 1 / \sqrt{1 + \delta_{j,j'}} $ for the two-body wave functions,
and the direct and exchange contributions are identical for $ k = \text{even} $,
and cancel for $ k = \text{odd} $.
\par
It is important to note that the Coulomb energy depends very much on the radial size of the wave function and, the latter might be different depending on the approach used to obtain it. This is the reason why we compare the r.m.s. radii of our HF calculations with experimental values in Table I of the main manuscript and obtain reasonable agreement (agreement within $ 0.3 \, \% $).
\section{Extra contributions to $ \Delta E $  in the Hartree-Fock (HF) calculation}
\subsection{Finite proton size}
\par
The charge distribution of the proton is parametrized by the form factor
$ f \left( \ve{r} \right) = \exp \left( - r^2/ a_p^2 \right) / \left( a_p \sqrt{\pi} \right)^3 $. 
Thus, the charge density is evaluated as
\begin{equation}
  \rho_c \left( \ve{r} \right)
  =
  \int
  d \ve{r}'
  \rho_p \left( \ve{r}' \right)
  f \left( \ve{r} - \ve{r}' \right).
\end{equation}
Here, we choose $ a_p = 0.65 \, \mathrm{fm} $
which corresponds to the r.m.s.~radius of the proton $ 0.8 \, \mathrm{fm} $,
a sufficient accuracy for our purposes.
\subsection{Center-of-mass effect}
\par
The center-of-mass correction is estimated by the harmonic-oscillator model;
the HF density is folded by the Gaussian function
$ g \left( \ve{r} \right) = \exp \left( r^2/ B^2 \right) / \left( B \sqrt{\pi} \right)^3 $
where $ B \equiv b/\sqrt{A} = \sqrt{\hbar/m\omega A} $.
The harmonic-oscillator frequency is given by
$ \hbar \omega = 45/A^{1/3} - 25/A^{2/3} \, \mathrm{MeV} $.
The calculated charge radii obtained from the HF wave functions of the core orbitals
reproduce well the empirical ones of
$ \nuc{O}{16}{} $ and $ \nuc{Ca}{40}{} $ in the both
parameter sets SGII and SAMi.
The calculated charge radii have good agreement with the experimental values
within $ 1 \, \% $ accuracy (see Table III of the main text).
This is an important constrain for realistic calculations of Coulomb energy,
whose value depends on the charge radii very much.
\subsection{Thomas-Ehrman effect and isospin impurity}
\par
In the mean-field calculations,
the proton wave function is different from the neutron one
in the same $ j $-orbital because of the Coulomb interaction,
i.e., the wave functions of valence protons are calculated by adding the Coulomb interaction
to the nuclear interaction adopted for the neutrons in the HF model.
This procedure is treated self-consistently in the HF calculations.
The self-consistent treatment gives rise to differences between the
nuclear potentials for protons and neutrons, i.e., the isospin impurity in the nuclear
potential.
These effects give rise to the following corrections to the binding energy,
\begin{align}
  \delta_{NN} \left( j \right)
  & =
    \brakket{\psi_j^p}{V_{NN}^p}{\psi_j^p}
    -
    \brakket{\psi_j^n}{V_{NN}^n}{\psi_j^n}
    \notag \\
  & =
    \left(
    \brakket{\psi_j^p}{V_{NN}^n}{\psi_j^p}
    -
    \brakket{\psi_j^n}{V_{NN}^n}{\psi_j^n}
    \right)
    +
    \brakket{\psi_j^p}{\left( V_{NN}^p - V_{NN}^n \right)}{\psi_j^p}
    \notag \\
  & \equiv
    \delta_{NN}^1 \left( j \right)
    +
    \delta_{NN}^2 \left( j \right),
\end{align}
where $ \psi_j^{\tau} $ and $ V_{NN}^{\tau} $ are the wave function of the $ j $-orbital and the mean-field potential for the nucleon $ \tau $ ($ \tau = p $, $ n $).
The term $ \delta_{NN}^1 \left( j \right) $
is due to the difference between the neutron and the proton wave functions of the valence orbital,
and is called the Thomas-Ehrman effect.
The term $ \delta_{NN}^2 \left( j \right) $
comes from the isospin impurity in the nuclear potential.
The correction $ \delta_{NN}^1 \left( j \right) $
appears in all mean-field calculations due to the Coulomb interaction for the protons,
while the $ \delta_{NN}^2 \left( j \right) $ term has a contribution only in the self-consistent Hartree or HF calculations.
\subsection{Electromagnetic spin-orbit interaction}
\par
The electromagnetic (EM) spin-orbit interaction is taken into account in the following form:
\begin{equation}
  V_{\urm{SO}}^{\urm{EM}}
  \left( \ve{r} \right)
  =
  \frac{\hbar^2}{4 m^2 c^2}
  \left[
    \left( g_p - 1 \right)
    \left( \frac{1}{2} - t_z \right)
    +
    g_n
    \left( \frac{1}{2} + t_z \right)
  \right]
  \times
  \left(
    \ve{l} \cdot \ve{\sigma}
  \right)
  \frac{1}{r}
  \frac{d V_{\urm{C}} \left( r \right)}{dr},
\end{equation}
where $ g_p = 5.58 $ and $ g_n = -3.82 $.
\subsection{Proton-neutron mass difference in the kinetic energy}  
\par
In the HF calculations, the kinetic energy is evaluated with the average mass of proton and neutron
$ \bar{m} = \left( m_p + m_n \right) / 2 $.  
The neutron and proton mass difference,
$ \Delta m = \left( m_n - m_p \right) $,
in the proton and neutron kinetic energies in the last orbital is taken into account by the following formula
with $ \bar{K}^{\urm{HF}} = \left( K_p^{\urm{HF}} + K_n^{\urm{HF}} \right) / 2 $ as
\begin{equation}
  K_p
  -
  K_n
  \simeq
  K_p^{\urm{HF}}
  \frac{\bar{m}}{m_p}
  -
  K_n^{\urm{HF}}
  \frac{\bar{m}}{m_n}
  \simeq 
  \bar{K}^{\urm{HF}}
  \frac{\Delta m}{\bar{m}}.
\end{equation}
\subsection{Core polarization}
\par
The core polarization energy caused by the valence nucleon is calculated by
performing constrained HF calculations for nuclei with mass
$ \left( A_{\urm{core}} + 1 \right) $.
Namely, the total binding energy
$ E_{\urm{HF}} \left( A_{\urm{core}} + 1 \right) $
is compared with the sum of the binding energy
$ E_{\urm{HF}} \left( A_{\urm{core}} \right) $
and the single-particle energy of the last nucleon $ \varepsilon \left( j \right) $
calculated in the core potential;
\begin{equation}
  \Delta E_{\urm{pol}}
  =
  E_{\urm{HF}} \left( A_{\urm{core}} + 1 \right)
  -
  \left[
    E_{\urm{HF}} \left( A_{\urm{core}} \right)
    +
    \varepsilon \left( j \right)
  \right].
\end{equation}
The correction on the energy difference between mirror nuclei is given by
\begin{equation}
  \delta_{\urm{pol}} \left( j \right)
  =
  \Delta E_{\urm{pol}}^p
  -
  \Delta E_{\urm{pol}}^n.
\end{equation}
The constrained HF calculations include a spurious self-energy contribution of
the last nucleon which should be subtracted. That is, the diagonal two-body matrix element for the last nucleon
$ E_{\urm{spur}} \left( i \right) \equiv \frac{1}{2} \brakket{ii}{\bar{V}}{ii} $
for $ i = A + 1 $ is included in the EDF automatically
[$ \bar{V} \left( \ve{r}_1, \ve{r}_2 \right) \equiv V \left( \ve{r}_1, \ve{r}_2 \right) \left( 1 - P_{12} \right) $
is the antisymmetrized interaction with $P_{12}$ the exchange operator between nucleons 1 and 2]. This is a spurious matrix element so that it is subtracted from the polarization energy as
$ E_{\urm{pol}} = E_{\urm{HF}} \left( A + 1 \right) - \left[ E_{\urm{HF}} \left( A \right) + \varepsilon \left( A + 1 \right) \right] - E_{\urm{spur}} \left( i \right) $.
\subsection{Vacuum polarization}
\par
The vacuum polarization effect due to the virtual emission and absorption of
electron-positron pairs gives
an additional repulsive potential between protons
as a lowest order correction in the fine structure constant $ \alpha \equiv e^2 / \hbar c $~\cite{
  Roca-Maza2018Phys.Rev.Lett.120_202501}.
The corresponding potential induced by the vacuum polarization can be written as~\cite{
  Uehling1935Phys.Rev.48_55}
\begin{equation}
  V_{\urm{VP}} \left( \ve{r} \right)
  =
  \frac{2}{3}
  \frac{\alpha e^2}{\pi}
  \int
  d \ve{r}'
  \frac{\rho_c \left( \ve{r}' \right)}{\left| \ve{r} - \ve{r}' \right|}
  \mathcal{K}_1 \left( \frac{2}{\lambdabar_e} \left| \ve{r} - \ve{r}' \right| \right)
\end{equation}
where $ \alpha $ is the fine-structure constant,
$ \lambdabar_e $ is the reduced Compton electron wavelength,
and
\begin{equation}
  \mathcal{K}_1 \left( x \right)
  =
  \int_1^{\infty}
  dt \,
  e^{-xt}
  \left( \frac{1}{t^2} + \frac{1}{2t^4} \right)
  \sqrt{t^2 - 1}.
\end{equation}
The vacuum polarization is estimated
to result in a $ 0.6 \, \% $ increase of the Coulomb energies of
both $ A = 17 $ and $ 41 $ systems.
\begingroup
\squeezetable
\begin{table}[tb]
  \centering
  \caption{
    Extra contributions to the mass difference of mirror nuclei of
    $ \left( N \pm 1, Z \right) $ and $ \left( N, Z \pm 1 \right) $ of
    $ A = N + Z = 16 $ and $ 40 $ closed shell core.
    The values are given in units of $ \mathrm{MeV} $.
    The core density and the wave function of valence orbital are calculated
    by HF model with the SGII Skyrme EDF.}
  \label{tab:tab-s1}
  \begin{ruledtabular}
    \begin{tabular}{ldddd}
      \multicolumn{1}{l}{Nuclei} & \multicolumn{1}{c}{$ \nuc{F}{17}{} $-$ \nuc{O}{17}{} $} & \multicolumn{1}{c}{$ \nuc{O}{15}{} $-$ \nuc{N}{15}{} $} & \multicolumn{1}{c}{$ \nuc{Sc}{41}{} $-$ \nuc{Ca}{41}{} $} & \multicolumn{1}{c}{$ \nuc{Ca}{39}{} $-$ \nuc{K}{39}{} $} \\
      \hline
      \multicolumn{1}{l}{Orbital} & \multicolumn{1}{c}{$ 1d_{5/2} $} & \multicolumn{1}{c}{$ \left( 1p_{1/2} \right)^{-1} $} & \multicolumn{1}{c}{$ 1f_{7/2} $} & \multicolumn{1}{c}{$ \left( 1d_{3/2} \right)^{-1} $} \\
      \hline
      Proton size effect                       & -0.053 & -0.070 & -0.066 & -0.082 \\
      Center-of-mass effect                    &  0.023 &  0.030 &  0.014 &  0.018 \\
      Thomas-Ehrman effect                     &  0.014 &  0.006 &  0.034 &  0.021 \\
      Isospin impurity                         &  0.050 & -0.136 &  0.134 & -0.176 \\
      EM Spin-orbit interaction                & -0.065 &  0.080 & -0.126 &  0.142 \\
      $ pn $ mass difference in kinetic energy &  0.034 &  0.024 &  0.040 &  0.031 \\
      Core polarization                        &  0.018 &  0.073 &  0.036 &  0.020 \\
      Vacuum polarization                      &  0.019 &  0.021 &  0.036 &  0.037 \\
      \hline
      Sum                                      &  0.040 &  0.028 &  0.102 &  0.011 \\
    \end{tabular}
  \end{ruledtabular}
\end{table}
\endgroup
\begingroup
\squeezetable
\begin{table}[tb]
  \centering
  \caption{
    The same as Table \ref{tab:tab-s1},
    but for the SAMi Skyrme EDF.}
  \label{tab:tab-s2}
  \begin{ruledtabular}
    \begin{tabular}{ldddd}
      \multicolumn{1}{l}{Nuclei} & \multicolumn{1}{c}{$ \nuc{F}{17}{} $-$ \nuc{O}{17}{} $} & \multicolumn{1}{c}{$ \nuc{O}{15}{} $-$ \nuc{N}{15}{} $} & \multicolumn{1}{c}{$ \nuc{Sc}{41}{} $-$ \nuc{Ca}{41}{} $} & \multicolumn{1}{c}{$ \nuc{Ca}{39}{} $-$ \nuc{K}{39}{} $} \\
      \hline
      \multicolumn{1}{l}{Orbital} & \multicolumn{1}{c}{$ 1d_{5/2} $} & \multicolumn{1}{c}{$ \left( 1p_{1/2} \right)^{-1} $} & \multicolumn{1}{c}{$ 1f_{7/2} $} & \multicolumn{1}{c}{$ \left( 1d_{3/2} \right)^{-1} $} \\
      \hline
      Proton size effect                       & -0.050 & -0.068 & -0.063 & -0.080 \\
      Center-of-mass effect                    &  0.021 &  0.029 &  0.014 &  0.018 \\
      Thomas-Ehrman effect                     &  0.014 &  0.007 &  0.031 &  0.021 \\
      Isospin impurity                         &  0.047 & -0.090 &  0.131 & -0.098 \\
      EM Spin-orbit interaction                & -0.061 &  0.078 & -0.121 &  0.140 \\
      $ pn $ mass difference in kinetic energy &  0.035 &  0.026 &  0.041 &  0.034 \\
      Core polarization                        &  0.018 &  0.073 &  0.036 &  0.020 \\
      Vacuum polarization                      &  0.019 &  0.020 &  0.035 &  0.037 \\
      \hline
      Sum                                      &  0.043 &  0.075 &  0.104 &  0.092 \\
    \end{tabular}
  \end{ruledtabular}
\end{table}
\endgroup
\subsection{Comparison between different approaches}
\par
In Table \ref{tab:CSB},
we compare the strengths of a Skyrme-like CSB interaction as estimated from different phenomenological ``Pheno''~\cite{
  Roca-Maza2018Phys.Rev.Lett.120_202501,
  Baczyk2018Phys.Lett.B778_178,
  Baczyk2019J.Phys.G46_03LT01,
  Sagawa2022Phys.Lett.B829_137072}
and microscopic ``Micro''~\cite{
  Novario2023Phys.Rev.Lett.130_032501,
  Wiringa}
approaches.
Within the latter, we also show the estimates presented in the main text of the manuscript for an easy comparison.
In the phenomenological approaches, a similar Skyrme-like interaction was used;
hence, we collect here the published values (accordingly rescaled) while we have used the approach detailed in Ref.~\cite{
  Naito2022Phys.Rev.C105_L021304}
to extract, in an approximate way, the equivalent Skyrme-like CSBI parameters of the microscopic calculations of Refs.~\cite{
  Novario2023Phys.Rev.Lett.130_032501,
  Wiringa}.
As it can be clearly seen from the table, depending on the theoretical method employed, the central values of the leading order term CSB parameter ($ \tilde{s}_0 $) can differ by an order of magnitude and could even be of different sign.
\begin{table}[tb]
  \centering
  \caption{Strengths of the various Skyrme-like CSB interactions. ``Pheno.''~and ``Micro.'',~respectively, refer to results based on phenomenological fits and on microscopic evaluation
    taken from Ref.~\cite{Naito2023}.
    See Ref.~\cite{Naito2023} and text for further details.}
  \label{tab:CSB}
  \begin{ruledtabular}
    \begin{tabular}{llllll}
      Class & Method or Name & \multicolumn{1}{c}{$ \tilde{s}_0 $ ($ \mathrm{MeV} \, \mathrm{fm}^3 $)} & \multicolumn{1}{c}{$ \tilde{s}_1 $ ($ \mathrm{MeV} \, \mathrm{fm}^5 $)} & \multicolumn{1}{c}{$ \tilde{s}_2 $ ($ \mathrm{MeV} \, \mathrm{fm}^5 $)} & Ref. \\
      \hline
      Pheno & SAMi-ISB                                          & $ -52.6 \pm  1.4 $ & \multicolumn{1}{c}{---} & \multicolumn{1}{c}{---} & \cite{Roca-Maza2018Phys.Rev.Lett.120_202501} \\
      Pheno & SLy4-ISB (leading order)                          & $ -22.4 \pm  4.4 $ & \multicolumn{1}{c}{---} & \multicolumn{1}{c}{---} & \cite{Baczyk2018Phys.Lett.B778_178} \\
      Pheno & SkM*-ISB (leading order)                          & $ -22.4 \pm  5.6 $ & \multicolumn{1}{c}{---} & \multicolumn{1}{c}{---} & \cite{Baczyk2018Phys.Lett.B778_178} \\
      Pheno & $ \text{SV}_{\text{T}} $-ISB (leading order)      & $ -29.6 \pm  7.6 $ & \multicolumn{1}{c}{---} & \multicolumn{1}{c}{---} & \cite{Baczyk2019J.Phys.G46_03LT01} \\
      Pheno & $ \text{SV}_{\text{T}} $-ISB (next-leading order) & $ +44   \pm  8   $ & $ -56 \pm 16 $ & $ -31.2 \pm 3.2 $ & \cite{Baczyk2019J.Phys.G46_03LT01} \\
      Pheno & Estimation by isovector density                   & $ -17.6 \pm 32.0 $ &  \multicolumn{1}{c}{---} & \multicolumn{1}{c}{---} & \cite{Sagawa2022Phys.Lett.B829_137072} \\
      \hline
      Micro & $ \Delta E_{\urm{tot}} $ ($ \text{N}^2 \text{LO}_{\text{GO}} $ (394) \& CC) & $ -4.2 \pm 6.5 $ &  \multicolumn{1}{c}{---} & \multicolumn{1}{c}{---} & \cite{Novario2023Phys.Rev.Lett.130_032501} \\
      Micro & $ \Delta E_{\urm{tot}} $ ($ \text{N}^2 \text{LO}_{\text{GO}} $ (450) \& CC) & $ -5.1 \pm 28.5 $ &  \multicolumn{1}{c}{---} & \multicolumn{1}{c}{---} & \cite{Novario2023Phys.Rev.Lett.130_032501} \\
      Micro & $ \Delta E_{\urm{tot}} $ (AV18-UX \& GFMC)                                  & $ -6.413 \pm 0.173 $ &  \multicolumn{1}{c}{---} & \multicolumn{1}{c}{---} & \cite{Wiringa} \\
      Micro & QCD sum rule (Case I)                                                       & $ -15.5^{+8.8}_{-12.5} $ & $ +0.52^{+0.42}_{-0.29} $ & \multicolumn{1}{c}{---} & Present \\
      Micro & QCD sum rule (Case II)                                                      & $ -15.5^{+8.8}_{-12.5} $ & \multicolumn{1}{c}{---} & $ +0.18^{+0.14}_{-0.10} $ & Present \\
    \end{tabular}
  \end{ruledtabular}
\end{table}